\documentstyle[aps,prl]{revtex}
\begin{document}
\bibliographystyle{prsty}
\draft

\title{First-principles computation of the Born effective charges \\
and their band-by-band decomposition}
\author{Ph. Ghosez and X. Gonze}
\address{Unit\'e de Physico-Chimie et de Physique des Mat\'eriaux,
Universit\'e Catholique de Louvain,\\
1 Place Croix du Sud, B-1348 Louvain-la-Neuve, Belgium}
\date{\today}
\maketitle
\begin{abstract}
The Born effective charge, $Z^*$, that describes the polarization
created by collective atomic displacements, can be computed
from first-principles by different techniques.
Its band-by-band decomposition appeared recently as a powerful tool
for the microscopic characterisation of the bonding in solids.
We describe the connections between the different expressions
used for the computation of $Z^*$, and
analyze the possible associated band-by-band decompositions. We show
that unlike for the full $Z^*$, the different band-by-band values are not
equal, and emphasize that one of them has an interesting physical meaning.
\end{abstract}

\setcounter{page}{1}

\section{Introduction}

The Born effective charge ($Z^*$) describes the macroscopic polarization
induced, in a crystalline insulating solid, by the collective displacements of
nuclei belonging to a given sublattice. In the lattice dynamics study of
insulating
crystals, this tensor is usually considered as a fundamental
quantity~\cite{Born33}, because it governs the amplitude of the long-range
Coulomb interaction between nuclei, and the splitting between longitudinal
(LO) and
transverse (TO) optic phonon modes.

In simple materials, like $A^N B^{8-N}$ binary crystals~\cite{Phillips70},
in which
phonon eigenvectors are imposed by symmetry, infra-red measurement of the
splitting between LO and TO modes allows an accurate estimation of $|Z^*|^2/
\epsilon_{\infty}$ and offers an unambiguous way to extract
the amplitude of $Z^*$ (its sign remains undefined)
from the experiment. However, in more complex materials like  ABO$_3$
compounds, LO and TO mode eigenvectors are not necessarily
equivalent and the determination of $Z^*$
from the experimental data is not so straightforward~: the value can only be
approximated within some ``realistic''  hypothesis~\cite{Axe67}.
For such compounds, the development of
theoretical methods giving directly access to $Z^*$ was therefore
particularly interesting.

Being defined in term of response to atomic displacements, the
Born effective charge is a {\it dynamical} concept that cannot be
assimilated to conventional static charges~\cite{Harrison80}.
In particular, the amplitude of $Z^*$ cannot
be estimated from the inspection  of the ground-state electronic density
alone~: It is also dependent on the electronic current flowing
through the crystal when nuclei are displaced.
In some materials, this current may be anomalously large and the amplitude of
$Z^*$ can deviate substantially from that of the nominal ionic
charges~\cite{Resta93,Ghosez94,Zhong94,Lee94b,Ogut96,Detraux97}.

During the early seventies, the physics of the Born effective charges was
already widely discussed within various semi-empirical
models~\cite{Harrison80,Lucovsky71,Lucovsky73,Lannoo73,Hubner75}. Since
that time, different theoretical advances have been performed
toward their first-principles determination.
A first powerful and systematic procedure was introduced by Baroni, Giannozzi
and Testa~\cite{Baroni87}, who suggested to determine $Z^*$ from a linear
response formalism grounded on a Sternheimer equation. A different algorithm,
based on a variational principle, was then reported by Gonze, Allan and
Teter~\cite{Gonze92}, yielding a new expression for $Z^*$. Thanks to recent
progress in the theory of the macroscopic polarization, $Z^*$ is now also
directly
accessible from finite difference of polarization~\cite{KingSmith93}. If
the first two methods were exclusively implemented within the density
functional formalism (DFT), the last one also allowed calculations of changes in
polarization within different other one-electron schemes (Hartree-Fock 
method~\cite{DallOlio97}, model GW approximations to many-body
theory~\cite{Massidda95,Gonze97c}, Harrison tight-binding
model~\cite{Bennetto96}) and the Hubbard tight-binding model~\cite{Resta95}.

The Born effective charges are now routinely computed within DFT and accurate
prediction have been reported for a large variety of materials. Without being
exhaustive, these calculations concern monoatomic crystals as
selenium~\cite{DalCorso93T}, various I-VII~\cite{Nardelli92,Wang94},
II-VI~\cite{Resta81,DalCorso93},  III-V~\cite{Giannozzi91,Bernardini97},
IV-IV~\cite{Karch94,Wang96a,Wellenhofer96} or IV-VI~\cite{Cockayne97}
semiconductors as well as a large diversity of oxydes
(AO~\cite{Schutt94,Posternak97}, AO$_2$~\cite{Gonze92,Lee94b,Detraux98},
AO$_3$~\cite{Detraux97},
ABO$_3$~\cite{Resta93,Ghosez94,Zhong94,Stixrude97}) and even more exotic
materials as Al$_2$Ru~\cite{Ogut96}. The previous results exclusively concern
crystalline materials but recently a calculation of $Z^*$ in a model of amorphous
silica has also been reported~\cite{Pasquarello97}.

Going beyond the bare determination of $Z^*$, the first-principles
approach was also offering a new opportunity to
investigate the microscopic mechanisms monitoring
its amplitude.  In some recent
studies~\cite{Ghosez95a,Ghosez95b,Massidda95,Posternak97}, the
decomposition of individual contributions from separate groups of occupied bands
to $Z^*$  appeared particularly
useful ~: The amplitude of
$Z^*$ can be related to dynamical changes of hybridizations and the band-by-band
decomposition allowed to identify the orbitals involved in such
a mechanism.

In the present paper, we aim at presenting the links between the theoretical
frameworks used nowadays for the computation of Born effective charges, and
deduce from this comparison the correct way to develop a band-by-band
analysis. We note that unitary transformations among occupied states,
called ``gauge tranformations'' are present at different levels
in the formalism. These unitary transforms need to be treated carefully
in order to obtain physical results. Also, the computation
of a band-by-band decomposition might be conceived from responses
to an homogeneous electric field or to a collective ionic displacement.
However, some care is needed here in order to yield physically meaningful
quantities.
The final expression that we propose is directly linked to displacements
of the center of gravity of Wannier functions.

In a first section, we make a brief overview of the
different procedures that, historically, have been considered
to determine $Z^*$ from first-principles.  In section II, we detail how $Z^*$
can be computed within DFT in the Berry phase approach and
the linear response formalism (either from
the response to an homogeneous electric field or a collective ionic
displacement).
We take into account the possible gauge transformations. Starting from the
general
expressions of $Z^*$, we then describe in Section III how contributions of
isolated
sets of bands can be separated from each others. We discuss the physical
significance of such band-by-band contributions in terms of Wannier functions.
We illustrate our discussion with a band-by-band decomposition of the titanium
charge in BaTiO$_3$, using different expressions.

\section{Overview of the different theoretical formulations of $Z^*$}

\subsection{Equivalent definitions of $Z^*$}

For periodic systems, the Born effective charge tensor
$Z^*_{\kappa,\alpha\beta}$ of nuclei belonging to the sublattice $\kappa$ is
conventionally defined as the coefficient of proportionality relating, under
the condition of zero macroscopic electric field, the change in
macroscopic polarization ${\cal P}_\beta$ along the
direction $\beta$ and the collective nuclear displacements of atoms
$\kappa$ along
direction $\alpha$, times the unit cell volume $\Omega_o$~:
\begin{equation}
\label{EqZ.P}
Z^*_{\kappa,\alpha\beta}
=
\Omega_0 \; 
  {\left .
   \frac{\partial{\cal P}_{\beta}}
           {\partial  \tau_{\kappa\alpha}} 
   \right |}_{{\cal E}=0}.
\end{equation}
We note the important condition of vanishing macroscopic electric field,
necessary to fix unambiguously the amplitude of the change of polarization.
Similarly, other dynamical charges may be defined from the change of
polarization
induced by a sublattice displacement when imposing other conditions on the field
${\cal E}$. These charges (Callen charge~\cite{Callen49}, Szigeti
charge~\cite{Szigeti49}) can be related to
$Z^*$ thanks to expressions involving the dielectric tensor (see, for instance,
Ref.~\cite{P2}).

The standard definition of $Z^*$, Eq.~(\ref{EqZ.P}), emphasizes the
response with respect to the collective nuclear displacement. However, a
thermodynamical equality relates the macroscopic polarization to the
derivative of
the electric enthalpy ${\tilde E}$ with respect to a homogeneous electric
field. Similarly another relationship  connects the forces on the nuclei to the
derivative of the electric enthalpy with respect to atomic displacements.
Combining these expressions, $Z^*$ can be  formulated, either as a mixed
second-order derivative of the electric enthalpy,
\begin{equation}
\label{EqZ.2}
Z^*_{\kappa,\alpha\beta}
=
- \frac{\partial^2 {\tilde E}}
           {\partial {\cal E}_\beta \partial \tau_{\kappa\alpha}},
\end{equation}
or as the derivative of the
force felt by a nucleus $\kappa$ with respect to an homogeneous effective
electric field ${\cal E}_\beta$, at zero atomic displacements~:
\begin{equation}
\label{EqZ.F}
Z^*_{\kappa,\alpha\beta}
=
    {\left .
    \frac{\partial F_{\kappa,\alpha}}
            {\partial {\cal E}_\beta}
    \right |}_{\tau_{\kappa \alpha}=0}.
\end{equation}
The three previous definitions -- Eqs.~(\ref{EqZ.P}),~(\ref{EqZ.2}), and
~(\ref{EqZ.F}) --
are formally equivalent. However, as it is now described,
each of them can lead to different algorithms for the computation of $Z^*$.

\subsection{The computation of $Z^*$ based on finite differences of the
polarization}

{}From Eq.~(\ref{EqZ.P}), it appears that an easy access to the macroscopic
polarization would allow the computation of $Z^*$ by finite difference. Starting
from the equilibrium positions of the atoms and considering a small but finite
collective displacement $\Delta \tau_{\kappa, \alpha}$ in direction
$\alpha$ of the
atoms belonging to the sublattice $\kappa$,
we obtain the following finite-difference estimate of $Z^*$ from the
change of polarization in zero-field~:
\begin{equation}
\label{dp}
Z^*_{\kappa,\alpha\beta}
=
\Omega_o \; \;
\lim_{\Delta \tau_{\kappa, \alpha} \rightarrow 0} \; \;
\frac{\Delta {\cal P}_{\beta}}{\Delta \tau_{\kappa, \alpha}}
\end{equation}

In the earliest first-principles computation of $Z^*$, in 1972,  Bennett and
Maradudin~\cite{Bennett72}  had attempted to deduce $Z^*$ using this technique,
but as pointed out by Martin~\cite{Martin74}, on the basis of an incorrect
expression for the polarization, yielding boundary-sensitive results. The basic
problem was that the change of polarization was assimilated to the change of the
unit cell dipole, which is ill-defined for delocalized periodic charge
distributions. A supercell technique, allowing correct estimation of the
polarization induced by longitudinal atomic displacements, was then proposed by
Martin and Kunc~\cite{Martin81,Kunc82} to compute the longitudinal charge,
giving indirectly access to $Z^*$ thanks to the knowledge of the optical
dielectric constant.

In 1992, Resta~\cite{Resta92} formulated the change of polarization as an
integrated macroscopic current. This yielded King-Smith and
Vanderbilt~\cite{KingSmith93} to identify in the change of electronic
polarization a geometric quantum phase (or Berry phase). They proposed
a new scheme, useful for a practical calculation of the macroscopic 
polarization~\cite{Resta94} and from which $Z^*$
can be deduced by finite difference, from single cell calculations.
It is usually referred to as the Berry phase approach.

\subsection{The computation of $Z^*$ based on linear responses}

Alternatively, the derivatives can also be computed explicitely using
perturbation theory. During the early seventies, the linear response of solids
to collective nuclear displacements was formulated in term of the inverse
dielectric matrix $\epsilon^{-1}$. Accordingly, an expression was proposed for
calculating the Born effective charges~\cite{Sham69,Pick70,Sham74}.
Computations based on this formalism were, for example, reported by Resta and
Baldereschi~\cite{Resta81,Baldereschi83}. However, one important drawback of
this procedure was the difficulty to control and to guarantee the charge
neutrality, which imposes constraints on the off-diagonal elements of
$\epsilon^{-1}$~\cite{Vogl78}. Consequently, Vogl~\cite{Vogl78} proposed a
method that bypasses the inversion of the dielectric function by using directly
the self-consistent potential induced by a long-wavelength lattice
displacement. Unfortunately, at that time, there was no way to determine
accurately this potential~: it had to be approximated and this formulation was
only applied to simplified models~\cite{Littlewood79a,Littlewood79b}. 

A solution to this problem was reported in 1987 by
Baroni, Giannozzi and Testa~\cite{Baroni87,Giannozzi91}  who proposed,
within  DFT, to compute the total effective potential by solving a
self-consistent
first-order Sternheimer equation. It was the first ``ab initio''
powerful and systematic approach yielding accurate
value of $Z^*$. Both types of perturbation
(the response to an homogeneous electric field
or to a collective ionic displacement) could be considered in this approach,
giving access to expressions linked to Eqs.~(\ref{EqZ.P}) and~(\ref{EqZ.F}).
A variational formulation of this theory was then reported
by Gonze, Allan
and Teter~\cite{Gonze92,Gonze97a,Gonze97b}, offering a different algorithm
for the
calculation of the
first-order wavefunctions and a new, stationary, expression for
$Z^*_{\kappa,\alpha\beta}$, linked to Eqs.~(\ref{EqZ.2}). The
``Sternheimer'' and
``variational'' formalisms were first
implemented within DFT when using plane-wave and different kind of
pseudopotentials~\cite{Giannozzi91,Ghosez94,Waghmare97,DalCorso97}
(usually within the LDA, but also within the GGA~\cite{DalCorso94}). Later,
LMTO~\cite{Savrasov92} and LAPW~\cite{Yu94} versions of the linear response
approaches have also been proposed.

\subsection{The computation of $Z^*$ based on finite differences of the atomic
forces}

In a similar spirit to the Berry phase approach, starting from Eq.
(\ref{EqZ.F}), $Z^*$
might also {\it a priori} be estimated from the force linearly  induced on the
different atoms by a finite homogeneous electric field, when keeping the atomic
positions frozen~:
\begin{equation}
Z^*_{\kappa,\alpha\beta}
=
\lim_{\Delta {\cal E}_{\beta} \rightarrow 0} \;
\frac{\Delta {F}_{\kappa, \alpha}}{\Delta {\cal E}_{\beta}}
\end{equation}
The force felt by an atom can  be trivially computed using the
Hellmann-Feynman theorem. However, the basic problem inherent
to this method is that infinite periodic systems (as those considered
in practical calculations) cannot sustain  a finite linear change
of potential. Different supercell techniques, that circumvent the problem,
have  been proposed to compute $Z^*$ by Kunc and Martin~\cite{Kunc82} (from the
force in the depolarizing field associated to a longitudinal phonon) and by Kunc
and Resta~\cite{Kunc83,Resta86} (from the force in an applied sawlike
potential). Their calculations demonstrate the feasability of the approach. Up to
now, it was however only marginally applied.  For that reason, in what follows,
we will concentrate on the linear response formalism and the Berry phase
approach.

\section{Analysis of the mathematical links between the different formulations}

\subsection{Ionic and electronic contributions}

For practical purposes, the Born effective charge can be conveniently
decomposed into two contributions~:
\begin{equation}
\label{Eq5.2}
Z^*_{\kappa,\alpha\beta}
= Z_\kappa \delta_{\alpha\beta} + Z_{\kappa,\alpha\beta}^{el}.
\end{equation}
The first term, $Z_\kappa$, is the charge of the nuclei (or
pseudo-ion, in case of
pseudopotential calculations), and can be trivially identified. The second,
$Z_{\kappa,\alpha\beta}^{el}$,
is the contribution due to the electrons, in response to the perturbation.
In this Section we analyse the mathematical links existing between
the different expressions that can be used to determine this last
contribution within the density functional formalism. Similar expressions should
be obtained within other frameworks as the Hartree-Fock method.

\subsection{The Berry phase approach}

As previously mentioned, a straightforward approach to the determination of
$Z^*_{\kappa}$ consists in computing the difference of macroscopic
polarization between a reference state, and a state where the atoms
belonging to the sublattice $\kappa$ have been displaced by a small but finite
distance $\Delta \tau_{\kappa, \alpha}$. Combining Eqs. (\ref{dp}) and
(\ref{Eq5.2}),
the electronic contribution to this charge can be obtained as~:
\begin{equation}
\label{EqPf}
Z^{el}_{\kappa,\alpha\beta}
=
\Omega_o
\lim_{\Delta \tau_{\kappa, \alpha} \rightarrow 0} \;
\frac{\Delta {\cal P}^{el}_{\beta}}{\Delta \tau_{\kappa, \alpha}}
\end{equation}
As demonstrated recently by King-Smith and
Vanderbilt~\cite{KingSmith93},  in periodic systems, the change in electronic
polarization in zero field can be deduced from the following formula~:
\begin{equation}
\label{Pol}
{\cal P}_{\beta}^{el}= - \frac{1}{(2 \pi)^3} \; i
   \sum_{n}^{occ} s
       \int_{BZ}
          \langle
   u_{n{\bf k}} | \frac{\partial}{\partial k_{\beta}} | u_{n{\bf k}}
          \rangle
      \; \;  d{\bf k}
\end{equation}
where $s$ is the occupation number of states in the valence bands ($s=2$ in
spin-degenerate system) and $u_{n{\bf k}}$ is the periodic part of the Bloch
functions. Taken independently at each wavevector {\bf k} in the Brillouin zone,
the matrix elements of the previous equation are ill-defined. Indeed, the phase of
the wavefunctions at each wavevector is arbitrary, and thus unrelated with the
phases at neighbouring wavevectors~: The derivative of the wavefunctions with
respect to their wavevector has a degree of arbitrariness.
However, the {\it integral} of the right-hand side is a
well-defined quantity, which has the
form of a Berry phase of band $n$, as discussed by Zak~\cite{Zak89}.

The King-Smith and Vanderbilt definition is valid  {\it only} if the wavefunctions
fulfill the {\em periodic gauge} condition. This means that the periodic part of
Bloch functions must satisfy
\begin{equation}
\label{Eq.PG}
u_{n {\bf k}}({\bf r})= e^{i {\bf G.r}} \;
u_{n {\bf k+G}}({\bf r}) ,
\end{equation}
This condition does not fix unambiguously the phase of the
wavefunctions at a given {\bf k}-point (even not at neighbouring {\bf k}-points)
but it imposes a constraint between wavefunctions at distant wavevectors. It
defines a topology in {\bf k}-space, within which the polarization takes the
convenient form of a Berry phase.

When working within one-electron schemes (DFT, Hartree-Fock, ...), a generalized
choice of phase is also present at {\it another} level.
For the ground-state, the Lagrange multiplier method
applied to the
minimization of the Hohenberg and Kohn fonctional under orthonormalization
conditions on
the wavefunctions~\cite{Gonze95b}, gives the following equations~:
\begin{equation}
\label{eq.s}
H_{\bf k} | u_{m{\bf k}} \rangle =
\sum_{n}^{occ} \Lambda_{mn,{\bf k}} \;
    | u_{n{\bf k}} \rangle
\end{equation}
This condition, associated with the minimisation of the Hohenberg
and Kohn energy functional, means that the wavefunctions $u_{n{\bf k}}$
must define a Hilbert space (right-hand side of Eq.~(\ref{eq.s})),
in which the vector generated by the application
of the Hamiltonian to one of these wavevectors (left-hand side of
Eq.~(\ref{eq.s})) must lie. We observe that a unitary transformation between the
wavefunctions will leave the Hilbert space invariant, and
Eq.~(\ref{eq.s}) will be satisfied provided the matrix of
Lagrange multiplier $\Lambda_{mn,{\bf k}}$ is transformed accordingly.
In order to build Kohn-Sham band structures, the unitary transform
is implicitely chosen such as to guarantee
\begin{equation}
\label{eq.diag}
\Lambda_{mn,{\bf k}}=
\delta_{mn} \; \epsilon_{m,{\bf k}}
\end{equation}
in which case
$\epsilon_{m,{\bf k}}$ correspond to the eigenvalues of the
Kohn-Sham Hamiltonian and the associated functions $u_{d,m{\bf k}}$ are the
Kohn-Sham orbitals. This choice is called the {\em diagonal gauge} condition. All
along this work, it will be emphasized by a ``$d$'' subscript.

We note that the periodic gauge condition connect wavefunctions at
different {\bf
k}-points, while the diagonal gauge condition fixes wavefunctions at a
given {\bf
k}-point. The choice defined by Eq.~(\ref{eq.diag}) is {\it not} mandatory, and the
computation of the total energy, the density, or the
Berry phase, Eq.~(\ref{Pol}), will give the {\it same} value independently
of the fulfillment of Eq.~(\ref{eq.diag}). If the diagonal gauge is
the natural choice for the ground-state wavefunctions, another choice is usually
preferred for the change in wavefunctions in linear-response calculations (see
below).

{}From a practical viewpoint, direct evaluation of Eq. (\ref{Pol}) is not
trivial in
numerical calculations because the wavefunctions are only computed  at a finite
number of points in the Brillouin zone, without any phase relationship
between the
eigenvectors. An elegant scheme to deal with this problem was reported in
Ref.~\cite{KingSmith93}.  Also, we note that, associated to the fact that a
phase is
only defined modulo $2
\pi$, Eq. (\ref{Pol}) only provides the polarization modulo a ``quantum'' (in
3-dimensional solids, the quantum is $(s . R_a / \Omega_0)$, where $R_a$ is a
vector of the reciprocal lattice). This quantum uncertainty should
appear as a limitation of the technique. However, for the purpose of
computing $Z^*$ from finite-differences,  the atomic
displacements $\Delta \tau_{\kappa, \alpha}$
may always be chosen sufficiently small for the associated change of
polarization being unambiguously defined.

The Berry phase technique was successfully applied to various ABO$_3$
compounds~\cite{Resta93,Zhong94}, giving results equivalent to those
obtained independently using
perturbative techniques~\cite{Ghosez94,Rabe94}. A similar agreement was for
instance reported for
alkaline-earth oxides~\cite{Schutt94,Posternak97}.

\subsection{$Z^*$ as the first derivative of the polarization}

Instead of approximating Eq. (\ref{EqZ.P}) from finite differences, it can
be chosen
to compute it directly.  The combination of Eqs.~(\ref{EqZ.P}),
(\ref{Eq5.2}) and
(\ref{Pol})  gives~:

\begin{eqnarray}
Z^{el}_{\kappa, \alpha \beta} =
 - \frac{\Omega_o}{(2 \pi)^3} \; i
   \sum_{m}^{occ} s
       \int_{BZ}
  [
          \langle
             \frac{\partial u_{n{\bf k}}}{\partial \tau_{\kappa, \alpha}}
 |
             \frac{\partial u_{n{\bf k}}}{\partial { k}_{\beta}}
          \rangle
 +
          \langle
            u_{n{\bf k}}
 |
            \frac{\partial}{\partial {k}_{\beta}}
 |
            \frac{\partial u_{n{\bf k}}}{\partial \tau_{\kappa, \alpha}}
          \rangle
 ]
       d{\bf k}.
\end{eqnarray}
The second expectation value can be worked out~:
\begin{eqnarray}
&&   \int_{BZ}
          \langle
            u_{n{\bf k}}
 |
            \frac{\partial}{\partial  { k}_{\beta}}
 |
             \frac{\partial u_{n{\bf k}}}{\partial \tau_{\kappa, \alpha}}
          \rangle
    d{\bf k}
=
\nonumber \\
&& \hspace{15mm}
     \int_{BZ}
 [
     \frac{\partial}{\partial  { k}_{\beta}}
          \langle
            u_{n{\bf k}}
 |
            \frac{\partial u_{n{\bf k}}}{\partial  \tau_{\kappa, \alpha}}
          \rangle
 -
          \langle
            \frac{\partial  u_{n{\bf k}}}{\partial  { k}_{\beta}}
 |
            \frac{\partial u_{n{\bf k}}}{\partial  \tau_{\kappa, \alpha}}
          \rangle
]
       d{\bf k}
\end{eqnarray}
In the previous expression, the first term of the right-hand side
is the gradient of a periodic quantity
integrated over the Brillouin zone. Within any periodic gauge,
Eq.~(\ref{Eq.PG}),
its contribution will be zero. Using the time-reversal symmetry,
we arrive therefore at the final expression~:
\begin{eqnarray}
\label{Z*B}
Z^{el}_{\kappa, \alpha \beta}
=  -2  \frac{\Omega_o}{(2 \pi)^3} \; i
   \sum_{n}^{occ}  s
       \int_{BZ}
       \langle
             \frac{\partial u_{n{\bf k}}}{\partial \tau_{\kappa, \alpha}}
 |
             \frac{\partial u_{n{\bf k}}}{\partial { k}_{\beta}}
      \rangle
      d{\bf k}
\end{eqnarray}

The first-derivatives of the wave functions, $\frac{\partial}{\partial
\tau_{\kappa,
\alpha}} u_{n{\bf k}}$ and
$\frac{\partial}{\partial {k}_{\beta}} u_{n{\bf k}}$, required to evaluate this
expression, can be computed by linear-response techniques either by
solving a first-order Sternheimer equation~\cite{Baroni87,Giannozzi91} or by
the direct minimization of a variational expression as described in
Ref.~\cite{Gonze92,Gonze97a}.

We note that the choice of gauge will influence the value of the
first-derivative of $u_{n {\bf k}}$, although the integrated quantity
$Z^{el}_{\kappa, \alpha \beta}$ must remain independent of this choice (in any
periodic gauge). Usually, the following choice is preferred in linear-response
calculations~:
\begin{eqnarray}
       \langle
            {\left .
            \frac{\partial u_{n{\bf k}}}{\partial \lambda}
            \right |}_p
 |
             u_{m{\bf k}}
      \rangle
 = 0
\end{eqnarray}
for $m$ and $n$ labelling occupied states, and $\lambda$
representing either the derivative with respect to the wavevector
or to atomic displacements. As emphasized by the notation (``$|_p$''), this
condition defines what is called the {\em parallel gauge} and insures that the
changes in the occupied wavefunctions are orthogonal to the space of the
ground-state occupied wavefunctions. This projection on the conduction bands is
not reproduced within the diagonal gauge defined by the generalization of
Eq.~(\ref{eq.diag}) at the first order of perturbation, as elaborated in
Ref.~\cite{Gonze95b}.

\subsection{$Z^*$ as a mixed second derivatrive of the electric enthalpy}

Considering now Eq. (\ref{EqZ.2}), $Z^*$ appears also as a mixed second
derivative of the  electric enthalpy. Therefore, following the formalism
used in Ref.~\cite{Gonze97a,Gonze97b}, $Z_{\kappa,\alpha\beta}^{el}$
can be formulated in terms of a {\it stationary} expression, involving the
first-order derivative of the wavefunctions with respect to a collective
displacement of atoms of the sublattice $\kappa$ 
and the first-order derivatives of the wavefunctions with respect to an electric
field and to their wavevector~\cite{Notation}~:

\begin{eqnarray}
\label{Eq5.3}
Z_{\kappa,\alpha\beta}^{el}
  &=&
 2\biggl[
   \frac {\Omega_0}{(2 \pi)^3}
\int_{\rm BZ}
 \sum_{m}^{\rm occ} s
   \biggl(
      \langle
		\frac{\partial u_{m{\bf k}}}{\partial \tau_{\kappa, \alpha}}
             |H_{\bf k}-\epsilon_{m{\bf k}}|
		\frac{\partial u_{m{\bf k}}}{\partial {\cal E}_\beta}
      \rangle
\nonumber \\
&&
 \,\,\, \,\,\,    + \langle
		\frac{\partial u_{m{\bf k}}}{\partial \tau_{\kappa, \alpha}}
   |   i
		\frac{\partial u_{m{\bf k}}}{\partial k_\beta}
       \rangle
      +\langle
        u_{m{\bf k}}
        |\frac{\partial v^{\prime}_{{\rm ext},{\bf k}}}{\partial
\tau_{\kappa \alpha}}|
		\frac{\partial u_{m{\bf k}}}{\partial {\cal E}_\beta}
       \rangle
   \biggr)  d{\bf k}
\nonumber \\
&&  \,\,\, + \frac{1}{2} \int_{\Omega_0}
    [\frac{\partial v_{\rm xc0}}{\partial  \tau_{\kappa\alpha}}({\bf r}) ]
    [\frac{\partial n}{\partial {\cal E}_\beta}({\bf r}) ]^*
   d{\bf r}
\nonumber \\
&&  \,\,\,   +2\pi\Omega_0
    \sum_{{\bf G}\neq 0}
    \frac{1} {|{\bf G}|^2}
    [\frac{ \partial {n}}   {\partial \tau_{\kappa \alpha}} ({\bf G}) ]^*
    \frac{ \partial n}{\partial {\cal E}_\beta}       ({\bf G})
\nonumber \\
&&  \,\,\,   +\frac{1}{2}  \int_{\Omega_0}
      K_{\rm xc}({\bf r,r'})
      [ \frac{ \partial {n}}    {\partial \tau_{\kappa \alpha}} ({\bf r}) ]^*
        \frac{ \partial n} {\partial {\cal E}_\beta}       ({\bf r'} )
       d{\bf r}  \; d{\bf r'} \,\,
\biggr]
\end{eqnarray}
\noindent

The {\em stationary} character of Eq.~(\ref{Eq5.3}) is related to the
influence of
inaccuracies of the derivatives
$\frac{\partial}{\partial \tau_{\kappa, \alpha}} u_{n{\bf k}}$
and
$\frac{\partial}{\partial {\cal E}_\beta} u_{n{\bf k}}$ on the accuracy of
$Z_{\kappa,\alpha\beta}^{el}$~: the error
in the latest is proportional to the {\it product} of the errors in the
two wavefunction derivatives. By contrast, 
Eq.~(\ref{Z*B}) does not have this property~: in that case, the errors on $Z^*$ is
directly proportional to the error on $\frac{\partial}{\partial \tau_{\kappa,
\alpha}} u_{n{\bf k}}$. As briefly discussed below, Eqs.~(\ref{Eq5.3}) and
(\ref{Z*B}) are mathematically equivalent. However,  if both are estimated with
approximate wavefunctions, Eq.~(\ref{Eq5.3}) is more accurate than
Eq.~(\ref{Z*B}).

Eq.~(\ref{Z*B}) can be seen to arise from the stationary property of
Eq.~(\ref{Eq5.3}). Indeed, in the latter, the error on $Z^*$ is proportional to the
product of the errors on the first-order change in wavefunctions so that if
$\frac{\partial}{\partial \tau_{\kappa, \alpha}} u_{n{\bf k}}$ was known
perfectly, a correct estimation of $Z_{\kappa,\alpha\beta}^{el}$ should be
obtained independently of the knowledge of $\frac{\partial}{\partial {\cal
E}_\beta} u_{n{\bf k}}$. Putting therefore to zero 
$\frac{\partial}{\partial {\cal E}_\beta} u_{n{\bf k}}$ and the corresponding
density changes in Eq.~(\ref{Eq5.3}), most of the terms cancel out and we recover
Eq. (\ref{Z*B}), which evaluated for the exact $\frac{\partial}{\partial
\tau_{\kappa, \alpha}} u_{n{\bf k}}$ must still correspond to a valid expression
for $Z^*$. Alternatively, the formal equivalence between Eqs.~(\ref{Z*B}) and
(\ref{Eq5.3}) should also be trivially established when introducing in Eq.
(\ref{Eq5.3}) the first-order Sternheimer equation associated to the atomic
displacement perturbation. 

\subsection{$Z^*$ as the first derivative of the atomic force}

By the same token as in the preceding Section, we can choose alternatively for
$\frac{\partial}{\partial \tau_{\kappa, \alpha}} u_{n{\bf k}}$ and
the associated density derivative to vanish
in Eq.~(\ref{Eq5.3}), and we still obtain a valid expression
for $Z^*$~:
\begin{eqnarray}
\label{Eq5.9}
Z_{\kappa, \alpha \beta}^{el}=
  &&
 2 \bigg[ \frac{\Omega_{0}}{(2 \pi)^3}
    \int_{BZ} \sum_{n}^{occ} s \;
       \langle
              u_{n{\bf k}}
              |
            \frac{\partial v_{{\rm ext},{\bf k}}^{\prime} }
                 {\partial \tau_{\kappa \alpha}           }
              |
            \frac{\partial u_{n{\bf k}}       }
                 {\partial {\cal E}_{\beta}   }
       \rangle
      d{\bf k}
\nonumber \\
&&  \,\,\,
+ \frac{1}{2}
  \int_{\Omega_0}
    [\frac {\partial v_{\rm xc0}          }
            {\partial  \tau_{\kappa\alpha} }
      ({\bf r}) ]
    [ \frac {\partial n          }
            {\partial {\cal E}_\beta  }
      ({\bf r}) ]^*
  d{\bf r}     \bigg]
\end{eqnarray}

This last equation corresponds to the third formulation of $Z^*$ in which it
appears as the first derivative of the force on the atoms $\kappa$ with
respect to an electric field (Eq. (\ref{EqZ.F})). Indeed, it is directly connected to
the following expression of the force, deduced from the Hellmann-Feynman
theorem~: 
\begin{eqnarray}
F_{\kappa, \alpha}^{el}=
  &&
  \frac{\Omega_{0}}{(2 \pi)^3}
    \int_{BZ} \sum_{n}^{occ} s \;
       \langle
              u_{n{\bf k}}
              |
            \frac{\partial v_{{\rm ext},{\bf k}}^{\prime} }
                 {\partial \tau_{\kappa \alpha}           }
              |
            u_{n{\bf k}}     
       \rangle
      d{\bf k}
\nonumber \\
&&  \,\,\,
+ 
  \int_{\Omega_0}
    [\frac {\partial v_{\rm xc0}          }
            {\partial  \tau_{\kappa\alpha} }
      ({\bf r}) ]
    [ n
      ({\bf r}) ]^*
  d{\bf r}     
\end{eqnarray}

Compared to Eqs. (\ref{Z*B}) and
(\ref{Eq5.3}), Eq. (\ref{Eq5.9}) has the advantage that the computation of the
first-order wavefunction derivative with respect to the electric field
perturbation is the {\it only} computationally intensive step needed to deduce
the {\it full} set of effective charges. 

We note however that the implementation of Eq. (\ref{Eq5.9}), rather easy within a
plane wave -- pseudopotential approach, is not so straightforward when the basis
set is dependent on the atomic positions, as in LAPW methods (additional Pulay
terms must be introduced). 

\section{A meaningful band-by-band decomposition}

\subsection{Problematics}

In the previous Section, we have described different approaches for
computing $Z^{el}_{\kappa, \alpha \beta}$. In order to have a better insight
on the mechanisms monitoring the amplitude of $Z^*$, it is
also useful to separate the partial contribution of electrons
from the different occupied bands to $Z^{el}_{\kappa, \alpha \beta}$. In what
follows, we describe how to identify the change of polarization induced by the
electrons of one specific band and we discuss the meaning of this
contribution in terms of Wannier functions.

\subsection{The displacement of center of gravity of Wannier functions}

Inspired by a previous discussion by Zak~\cite{Zak89}, Vanderbilt and
King-Smith~\cite{Vanderbilt93} emphasized that the macroscopic electronic
polarization acquires a particular meaning when expressed in terms of localized
Wannier functions. The periodic part of
Bloch functions  $u_{n{\bf k}}({\bf r})$
are related to the Wannier
functions $W_n({\bf r})$ through
the following transformation~:
\begin{eqnarray}
\label{eq.uW}
 u_{n{\bf k}}({\bf r}) &=& \frac{1}{\sqrt{N}}
       \sum_{\bf R} \; e^{-i{\bf k.(r-R)}} \; W_n{\bf (r-R)} \\
\end{eqnarray}
\begin{eqnarray}
\label{Eq.Wu}
 W_n{\bf (r)} &=& \frac{\sqrt{N} \; \Omega_o}{(2 \pi)^3}
       \int_{BZ} \; e^{i{\bf k.r}} \; u_{n{\bf k}}({\bf r}) \;
d{\bf k}
\end{eqnarray}
{}From this definition, we deduce that~:
\begin{eqnarray}
\frac{\partial}{\partial {\bf k}_{\beta}}  u_{n{\bf k}}({\bf r})
=
\frac{1}{\sqrt{N}} \;
       \sum_{R} \;\; [-i (r_{\beta}-R_{\beta})] \;
                            e^{-i{\bf k.(r-R)}} \; W_n({\bf r-R})
\end{eqnarray}
where {\bf R} runs over all real space lattice vectors. Introducing this
result in the equation providing $P_{\beta}$ (Eq. \ref{Pol}), we obtain~:
\begin{eqnarray}
{\cal P}_{\beta}^{el} = \frac{s}{\Omega_o}
     \sum_{n}^{occ}
       \int  r_{\beta} . | W_n({\bf r}) |^2 \;d{\bf r}
\label{eq.PtW}
\end{eqnarray}
{}From this equation, the electronic part of the polarization is simply
deduced from the position of the center of gravity of the electronic charge
distribution, as expressed in terms of localized Wannier functions. In other
words, for the purpose of determining the polarization, {\it ``the true
quantum mechanical electronic system can be considered as an effective
classical system of quantized point charges, located at the centers of
gravity associated with the occupied Wannier functions in each unit
cell''}~\cite{Vanderbilt93}.

We observe that Eqs. (\ref{eq.uW}) and (\ref{Eq.Wu}) establish a one-to-one
correspondence between $u_{n {\bf k}}$ and $W_n$.  As previously
emphasized in Section III, when working within the {\it diagonal} gauge,
$u_{d,n {\bf k}}$ becomes identified with the Kohn-Sham orbitals so that the
associated $W_{d,n}$ will correspond to a single band Wannier function.
Within this specific gauge, we can therefore isolate $P_{m,\beta}$, the
contribution of band $m$ to the $\beta$ component of the polarization,
by separating the different term in the sum appearing in Eq. (\ref{eq.PtW})~:
\begin{eqnarray}
{\cal P}_{m,\beta}^{el}= \frac{s}{\Omega_o}
       \int  r_{\beta} . | W_d,m({\bf r}) |^2 d{\bf r}
\end{eqnarray}

If we take now the derivative of the polarization with respect to a collective
atomic displacement, $Z^{el}_{\kappa, \alpha \beta}$ can be written in terms
of Wannier functions as~:
\begin{eqnarray}
\label{Z*W}
Z^{el}_{\kappa, \alpha \beta} =
  \sum_{n}^{occ}   s
     \int  r_{\beta}
         [ ({\left . \frac{\partial W_n({\bf r})}{\partial \tau_{\kappa,
\alpha}} \right |}_{d} )^*
             \;  W_{d,n}({\bf r})
         +
           (W_{d,n}({\bf r}))^*
             \;  {\left . \frac{\partial W_n({\bf r})}{\partial \tau_{\kappa,
\alpha}}\right |}_{d}
         ]
      d{\bf r}
\end{eqnarray}
As for the polarization, this equation has a simple physical meaning. In
response to an atomic displacement, the electronic distribution is modified
and the electronic contribution to $Z^*$ can be identified from the
displacement of the center of gravity of the occupied Wannier functions.
Working within the diagonal gauge at any order of perturbation, we will be
able to follow the change of single band Wannier functions all along the path of
atomic displacements. In the previous
expression,  the contribution of band
$m$ to
$Z^{el}_{\kappa, \alpha \beta}$ can therefore also be isolated~:
\begin{eqnarray}
\label{Z*Wn}
[Z^{el}_{\kappa, \alpha \beta}]_m =
   s  \int  r_{\beta}
         [ ({\left . \frac{\partial W_m({\bf r})}{\partial \tau_{\kappa,
\alpha}} \right |}_{d})^*
             \;  W_{d,m}({\bf r})
         +
           (W_{d,m}({\bf r}))^*
             \;  {\left . \frac{\partial W_m({\bf r})}{\partial \tau_{\kappa,
\alpha}} \right |}_{d}
         ]
      d{\bf r}
\end{eqnarray}
This last equation identifies the contribution from band $m$ to the Born
effective charge as $\Omega_0$ times the {\it change of polarization
corresponding to the displacement of a point charge $s$ on a distance equal to the
displacement of the Wannier center of this band}.

Eq.~(\ref{Z*Wn}) can also easily be evaluated in reciprocal space~:
\begin{eqnarray}
\label{Z*Bn}
[Z^{el}_{\kappa, \alpha \beta}]_m
=  -2 \frac{\Omega_o}{(2 \pi)^3} \; i \; s
       \int_{BZ}
       \langle
            {\left . 
            \frac{\partial u_{m{\bf k}}}{\partial \tau_{\kappa, \alpha}}
            \right |}_{d}
 |
            {\left . 
             \frac{\partial u_{m{\bf k}}}{\partial { k}_{\beta}}
            \right |}_{d}
      \rangle
      d{\bf k}
\end{eqnarray}
As Bloch and Wannier functions were related through a band-by-band
transformation, the contribution from band $m$ to $Z^*_{\kappa, \alpha
\beta}$ in Eq. (\ref{Z*Bn}) keeps the same clear physical meaning as in Eq.
(\ref{Z*Wn})~: $[Z^{el}_{\kappa, \alpha \beta}]_m$ corresponds to  $\Omega_0
\; . \; \Delta{\cal P}^{el}_{m,\beta}= \Omega_0 \; (\; s . \Delta d_{\beta})$ where
$\Delta d_{\beta}$ is the displacement in direction $\beta$ of the  Wannier
center of band $m$ induced by the unitary displacement of the sublattice of
atoms $\kappa$ in  direction $\alpha$.
In practical calculations, where each band may be thought as a
combinaison of well-known orbitals, the displacement of the Wannier center
is associated to the admixture of a new orbital character to the band and must be
attributed to dynamical changes of orbital hybridizations. As illustrated
in some recent studies~\cite{Ghosez95a,Ghosez95b,Massidda95,Posternak97}, the
decomposition of $Z^*$ appears therefore as a powerful tool for the microscopic
characterisation of the bonding in solids.

Let us emphasize again that the previous decomposition in terms of a single band
is valid only if the {\it diagonal} gauge was used to define the Kohn-Sham
wavefunctions, hence the ``$d$'' subscript in Eq.~(\ref{Z*Wn}) and (\ref{Z*Bn}).
The ground-state wavefunctions are conventionally computed within the diagonal
gauge. However, in most calculations,  the first-derivatives of these
wavefunctions are computed within the {\it parallel} gauge. Within this choice,
the change in each Bloch function will be a mixing of different Kohn-Sham
orbitals when the perturbation is applied so that the associated change in
functions $W_n$, defined from Eq. (\ref{Eq.Wu}), will correspond to the change
of a multi-band Wannier function. Evaluating Eq. (\ref{Z*Wn}) or (\ref{Z*Bn})
within such a gauge, we will identify the displacement of a complex of bands but
not of a single band. In practice, the first-order derivative of wavefunctions in
the diagonal gauge $\frac {d u_{n{\bf k}}}{d \lambda} |_d$ can be deduced from
those in the parallel gauge $\frac {d u_{n{\bf k}}}{d \lambda} |_p$ and the
ground-state wavefunctions in the diagonal gauge $u_{d,n{\bf k}}$, by adding
contributions from the subspace of the occupied bands~:
\begin{equation}
\label{Eq.gaugechange}
{\left . \frac {d u_{m{\bf k}}}{d \lambda} \right |}_d  =
{\left . \frac {d u_{m{\bf k}}}{d \lambda} \right |}_p
- \sum_{n \neq m}^{occ}
    \frac{   \langle u_{d,n{\bf k}}
               |  \frac{\partial H} {\partial \lambda}
               |  u_{d,m{\bf k}}
             \rangle
          }
 {   (\epsilon_{n{\bf k}} - \epsilon_{m{\bf k}})
          }
    u_{d,n{\bf k}}
\end{equation}
We note that this transformation (Eq.~(\ref{Eq.gaugechange})) can present some
problems when the denominator vanishes~: this happens when the valence energies
are degenerated. The problem can be partly bypassed by keeping a parallel
transport gauge within the space of degenerated wavefunctions. Practically, this
means that we will only be able to separate the contributions of disconnected
set of bands.

\subsection{Other band-by-band decompositions}

Up to now, we focused on Eq.~(\ref{Z*B}). We would like to investigate
whether
other band-by-band decompositions could be obtained from Eqs.~(\ref{Eq5.3})
and~(\ref{Eq5.9}). These expressions, unlike Eq.~(\ref{Z*B}), are not
written as a
simple sum of matrix elements, each related with a single band. However,
Eq.~(\ref{Eq5.9}) can still be transformed using a decomposition of the density
with respect to the bands~:
\begin{eqnarray}
\label{Eq.n}
n({\bf r})=
  &&
\frac{1}{(2 \pi)^3}
    \int_{BZ} \sum_{n}^{occ} s \;
              u_{n{\bf k}}^*({\bf r})
              u_{n{\bf k}}({\bf r})
      d{\bf k}.
\end{eqnarray}
It gives~:
\begin{eqnarray}
\label{Eq5.9bis}
Z_{\kappa, \alpha \beta}^{el}=
  &&
 2 \frac{\Omega_{0}}{(2 \pi)^3}
    \int_{BZ} \sum_{n}^{occ} s \;
       \langle
              u_{n{\bf k}}
              |
            \frac{\partial v_{{\rm ext},{\bf k}}^{\prime} }
                 {\partial \tau_{\kappa \alpha}           }
             +
												\frac {\partial v_{\rm xc0}         }
            					{\partial  \tau_{\kappa\alpha} }
              |
            \frac{\partial u_{n{\bf k}}       }
                 {\partial {\cal E}_{\beta}   }
       \rangle
      d{\bf k}
\end{eqnarray}
for which the following decomposition is obtained, using
the diagonal gauge wavefunctions~:
\begin{eqnarray}
\label{Eq5.9Bn}
[\tilde Z_{\kappa, \alpha \beta}^{el}]_m=
  &&
 2 \frac{\Omega_{0}}{(2 \pi)^3}
    \int_{BZ} s \;
       \langle
              u_{d,m{\bf k}}
              |
            \frac{\partial v_{{\rm ext},{\bf k}}^{\prime} }
                 {\partial \tau_{\kappa \alpha}           }
             +
												\frac {\partial v_{\rm xc0}         }
            					{\partial  \tau_{\kappa\alpha} }
              |
            {\left .  \frac{\partial u_{m{\bf k}}       }
                 {\partial {\cal E}_{\beta}   }  \right |}_d    
       \rangle
      d{\bf k}
\end{eqnarray}
This expression corresponds to the contribution of the electrons of
band $m$ to the force induced on atom $\kappa$ by a macroscopic field
${\cal E}_{\beta}$. However, it is not equivalent to Eqs.~(\ref{Z*Wn})
or ~(\ref{Z*Bn}). Indeed, for a particular band
$m$, the difference between matrix elements present in Eq.~(\ref{Z*Bn})
and~(\ref{Eq5.9Bn}) is (within a given gauge)~:
\begin{eqnarray}
\label{Eq.diff}
&& \hspace{-20mm}
\bigg[
   \langle
              u_{m{\bf k}}
              |
            \frac{\partial v_{{\rm ext},{\bf k}}^{\prime} }
                 {\partial \tau_{\kappa \alpha}           }
      +
												\frac {\partial v_{\rm xc0}         }
            					{\partial  \tau_{\kappa\alpha} }
              |
            \frac{\partial u_{m{\bf k}}       }
                 {\partial {\cal E}_{\beta}   }
   \rangle
\bigg]
-
\bigg[
     \langle
              \frac{ \partial u_{m{\bf k}}           }
                   { \partial \tau_{\kappa, \alpha}  }
              |
              -i \frac{ \partial u_{m{\bf k}}    }
                      { \partial {k}_{\beta}     }
     \rangle
\bigg]
\nonumber \\
& & \hspace{15mm}
-
 \frac{1}{2}  \int_{\Omega_0}
      K_{\rm xc}({\bf r,r'})
      [ \frac{ \partial {n}                 }
             { \partial \tau_{\kappa \alpha}}
        ({\bf r}) ]^*
        \frac{ \partial n_{m{\bf k}}     }
             { \partial {\cal E}_\beta   }
        ({\bf r'} )
       d{\bf r}  \; d{\bf r'}
\nonumber \\
 & &  \hspace{15mm}
+
 \frac{1}{2}  \int_{\Omega_0}
      K_{\rm xc}({\bf r,r'})
      [ \frac{ \partial n_{m{\bf k}}         }
             { \partial \tau_{\kappa \alpha} }
        ({\bf r})
                 ]^*
        \frac{ \partial n                    }
             { \partial {\cal E}_\beta       }
        ({\bf r'} )
       d{\bf r}  \; d{\bf r'}
\end{eqnarray}
where $n_{m{\bf k}}({\bf r})$ is a short notation for $u_{m{\bf k}}^*({\bf r})
u_{m{\bf k}}({\bf r})$. The summation of these differences on all the bands and
integration on the Brillouin zone gives zero, as expected. However, the
band-by-band difference, Eq.~(\ref{Eq.diff}), does not vanish. The quantity defined
from Eq. (\ref{Eq5.9Bn}) is therefore independent from that of Eq.~(\ref{Z*Wn})
and has no specific meaning in terms of Wannier functions.

\subsection{Numerical comparison of the different decompositions}

The previous theoretical results can now be illustrated on a numerical
example. In what follows, we will consider the case of barium titanate, a
well-known ferroelectric material of the ABO$_3$ perovskite family, presenting
non-trivial values of $Z^*$. Our calculations have been performed within a
planewave-pseudopotential approach. The electronic wavefunctions have been
expanded in plane-waves up to a kinetic energy cutoff of 35 hartrees. Integrals
over the Brillouin zone have been replaced by sums on a $6 \times 6 \times 6$
mesh of special k-points. The Born effective charges have been obtained by linear
response. 

In Table~\ref{Zbbb}, we compare different decompositions of the
titanium charge ($Z^*_{Ti}$), in the cubic phase. We observe
that the results obtained within the diagonal and parallel gauge, either from
Eq.~(\ref{Z*Bn}) or ~(\ref{Eq5.9Bn}) are {\it significantly} different. As
demonstrated previously, the identification of meaningful band-by-band
contributions require the use of Eq.~(\ref{Z*Bn}), when working within the
diagonal gauge~:  the contributions then describe the displacement of the
Wannier center of each given set of bands, induced in response to the
displacement of the Ti atom. A similar decomposition of the Born effective
charge for the other atoms in the unit cell is reported in
Refs.~\cite{Ghosez95b,P2}. In these papers, the origin of anomalous contributions
(i.e. the deviation with respect to the reference ionic values) is discussed in
terms of dynamical changes of orbital hybridization.

\section{Conclusions}

In this paper, we have recalled the three equivalent definitions of
$Z^*$ and we have discussed the associated mathematical formulations
within the density functional formalism.  A particular attention has been
placed on the gauge freedom associated to unitary transforms within the
subspace of the occupied bands.

Eq. (\ref{Eq5.3}) expresses $Z^*$ as a second derivative of the electric enthalpy.
Contrary to the other expressions, it has a stationary character and allows the
most accurate estimate of $Z^*$ when approximate wavefunctions are used. 

Eq. (\ref{Eq5.9}) formulates $Z^*$ as the force linearly induced on atoms $\kappa$
when a macroscopic electric is applied. It is an alternative convenient expression
from which the full set of effective charges can be deduced as soon as the
derivative of the wavefunctions with respect to the electric field perturbation is
known. 

Eq. (\ref{Z*B}) and its finite difference expression -- Eq.(\ref{EqPf}) --
consider $Z^*$ in terms of the macroscopic polarization induced by the
displacement of the atoms belonging to a given sublattice. Contrary to the other
formulations, it yields a meaningful band-by-band decomposition, helpful in the
characterisation of the bonding in solids.  It has been demonstrated that the
contribution of a particular band $m$ to $Z^*_{\kappa,\alpha \beta}$, obtained
from this expression when working within the diagonal gauge, is directly related
to the displacement of the Wannier center of this band when a sublattice of
atoms is displaced.

Finally, it has been shown explicitely, for BaTiO$_3$, that the
band-by-band decompositions, arising from Eqs. (\ref{Z*B}), (\ref{Eq5.3}) or
(\ref{Eq5.9}), within the parallel or diagonal gauge, yield significantly different
results.


\begin{table}
\caption {Band-by-band decompositions of the Born effective charge of
the Ti atom in the cubic phase of BaTiO$_3$. In the second column are mentioned
reference values expected in a purely ionic material; band-by-band contributions
presented in the three next columns were deduced from first-principles
calculations. Only the values obtained from Eq.~(\protect\ref{Z*Bn})
and within the diagonal gauge can be understood in terms of Wannier functions.}
\label{Zbbb}
\begin{center}
\begin {tabular}{lcccc}
&Reference  &\multicolumn{2}{c}{Diagonal gauge}
&Parallel gauge \\
& ionic values  & from Eq. (\ref{Z*Bn}) & from Eq. (\ref{Eq5.9Bn})
& from Eq. (\ref{Z*Bn})\\
\hline
core    &$+12.00$        &$+12.00$      &$+12.00$      &$+12.00$ \\
Ti 3s   &$-2.00$     &$-2.03$   &$+1.56$   &$-0.36$\\
Ti 3p   &$-6.00$     &$-6.22$   &$-9.54$   &$-5.50$\\
Ba 5s   &$0.00$      &$+0.05$   &$-0.36$   &$0.00$\\
O2s     &$0.00$     &$+0.23$   &$-1.56$   &$-0.41$\\
Ba 5p   &$0.00$     &$+0.36$   &$+1.47$   &$+0.10$\\
O 2p    &$0.00$     &$+2.86$   &$+3.68$   &$+1.42$\\
\hline
$Z^*_{Ti}$  &$+4.00$    &$+7.25$    &$+7.25$  &$+7.25$\\
\end{tabular}
\end{center}
\end{table}


\begin{thebibliography}{10}
\bibitem{Notation}
The notation slightly differs from Ref.~\cite{Gonze97a,Gonze97b}~:
the derivative are explicitely written, the ${\bf q=0}$ character
of the perturbation is omitted. Note that the ``prime'' indicates that the ${\bf
G}=0$ term has been omitted.
\end{thebibliography}
\end{document}